\documentclass[a4paper]{IEEEtran}
\IEEEoverridecommandlockouts
\usepackage{cite}
\usepackage{amsthm}
\usepackage{amsmath,amssymb,amsfonts}
\usepackage{algorithm}
\usepackage{algorithmicx}
\usepackage{algpseudocode}
\usepackage{geometry}
\usepackage{graphicx}
\usepackage{textcomp}
\usepackage{xcolor}
\usepackage{bm}
\usepackage{graphicx,epstopdf,psfrag,url,cite,xcolor,multirow,float}
\usepackage[font=footnotesize]{subfig}

\graphicspath{{figures/}}
\usepackage{times}
\begin{document}
\title{Joint Traffic Reshaping and Channel Reconfiguration in RIS-assisted Semantic NOMA Communications}
\author{Songhan Zhao\IEEEauthorrefmark{1}, Yusi Long\IEEEauthorrefmark{1}, Lanhua Li\IEEEauthorrefmark{1}, Bo Gu\IEEEauthorrefmark{1}, Shimin Gong\IEEEauthorrefmark{1}, and Zehui Xiong\IEEEauthorrefmark{3}\\
\IEEEauthorblockA{
\IEEEauthorrefmark{1}School of Intelligent Systems Engineering, Sun Yat-sen University, China\\
\IEEEauthorrefmark{3} School of Electronics, Electrical Engineering and Computer Science, Queen's University Belfast, U.K. \\
}

\vspace{-1.1cm}
}
\maketitle
\thispagestyle{empty}
\begin{abstract}
In this paper, we consider a semantic-aware reconfigurable intelligent surface (RIS)-assisted wireless network, where multiple semantic users (SUs) simultaneously transmit semantic information to an access point (AP) by using the non-orthogonal multiple access (NOMA) method. The SUs can reshape their traffic demands by modifying the semantic extraction factor, while the RIS can reconfigure the channel conditions via the passive beamforming. This provides the AP with greater flexibility to decode the superimposed signals from the SUs. We aim to minimize the system's overall energy consumption, while ensuring that each SU's traffic demand is satisfied. Hence, we formulate a joint optimization problem of the SUs' decoding order and semantic control, as well as the RIS's passive beamforming strategy. This problem is intractable due to the complicated coupling in constraints. To solve this, we decompose the original problem into two subproblems and solve them by using a series of approximate methods. Numerical results show that the joint traffic reshaping and channel reconfiguration scheme significantly improves the energy saving performance of the NOMA transmissions compared to the benchmark methods.
\end{abstract}
\vspace{-0.2cm}
\section{Introduction}

With the development of the sixth-generation (6G) wireless networks, the rapid growth of internet of things (IoT) and their increasing traffic demands pose significant challenges for numerous connections~\cite{Ding-2022pro}. Conventional multiple access methods, e.g., the time/frequency division multiple access (TDMA/FDMA), serve the individual IoT device in the orthogonal wireless resource block aiming to avoid co-channel interference. However, these methods are facing intensely competitive spectral utilization, which makes them challenging to meet the high demands in the expanding IoT ecosystem~\cite{Yuan-2021icm}. The non-orthogonal multiple access (NOMA) technique has been recognized as a promising solution to achieve higher multiplexing gain~\cite{Liu-2022jasc}. The NOMA technique enables multiple users to simultaneously access the wireless channel by sharing the same wireless resource block. The access point (AP) decodes each NOMA user's signal from the superimposed signals by employing the successive interference cancellation (SIC) method. However, the successful decoding  requires a sufficient spatial orthogonality among users, which is challenging to achieve in dense networks.

The reconfigurable intelligent surface (RIS) as a promising technique is expected to solve the spatial orthogonality problem in NOMA wireless networks.  The RIS is equipped with massive passive elements, which can reconfigure the propagation environment on demand by inducing phase shifts on the incident signals~\cite{Gong-2020cst}. 
Motivated by this, the authors in~\cite{Liu-2021jasc} maximized the energy efficiency in the RIS-assisted NOMA wireless networks by jointly optimizing the RIS's deployment and passive beamforming. This research indicates that the reasonable deployment of the RIS is crucial for enhancing the NOMA transmission performance. The authors in~\cite{Wang-2022twc} investigated the impact of the RIS on improving the NOMA transmission security, in which the RIS has been validated to improve the NOMA transmissions while preventing signal leakage to the eavesdroppers. Besides the fixed-location RIS, the RIS can be also mounted on the unmanned aerial vehicles (UAVs) as the aerial-RIS to provide more flexible reflection services, as explored in~\cite{Zhang-2023twc}. By jointly optimizing the trajectory and passive beamforming, the aerial-RIS can meet the dynamic traffic demands more effectively. However, even with the channel gain improved by the RIS, the NOMA users may still struggle to work efficiently under the heavy traffic demands. 

Recently, the semantic communication has emerged as another promising technique introducing a new control dimension to wireless networks. Different from the bit-oriented communications by the Shannon's information theory, the semantic communication focuses on the content-oriented communications by transmitting the key semantic information extracted from the raw data~\cite{Deniz-2023jasc}. By controlling the depth of semantic extraction, the size of the transmitted information can be compressed, and thus allowing for the on-demand reshaping of the users' traffic demands~\cite{Yang-2023jasc}. The authors in~\cite{Mu-2023jsac} investigated employing the semantic communication to a two-user NOMA system, where the primary user employs conventional bit communication while the secondary user transmits semantic information. Their findings validate that employing the semantic communication for the secondary user can improve the overall NOMA transmission performance without compromising the primary user's performance. Note that the achieved rate of the NOMA users varies significantly depending on their decoding order. Typically, the NOMA decoding order can be determined based on the users' channel conditions. We can further employ the RIS to make the channel conditions more controllable, providing greater flexibility in designing the NOMA decoding order. However, as the number of users increases, the ``early-late" issue will become more prominent~\cite{Mu-2023jsac}. The users decoded earlier suffer stronger interference from those decoded later, which may prevent them from meeting the traffic demands. Another issue that should be considered is how to balance the energy trade-off between the semantic control and transmission. Although the semantic extraction reduces energy consumption in transmission, it requires additional energy for the semantic control at both the AP and the user side.

To tackle the above difficulties,  we explore a joint traffic reshaping and channel reconfiguration scheme for the NOMA transmissions. In particular, the RIS's passive beamforming reconfigures the channel conditions while the semantic control reshapes the traffic demands of the semantic users (SUs), which are expected to improve the SUs' NOMA transmissions more efficiently. Meanwhile, this joint scheme also enhances the flexibility in managing the SUs' decoding order. We minimize the system's overall energy consumption by jointly optimizing the SUs' decoding order and semantic control, as well as the RIS's passive beamforming strategy. The complex problem is decomposed into two subproblems, each of which is solved by using either the penalty relaxation or the quadratic transform (QT) method. Numerical results verify the joint traffic reshaping and channel reconfiguration scheme significantly improves energy saving performance. The performance gain is further enhanced when the RIS is deployed closer to either the AP or the SUs.
\vspace{-0.1cm}
\section{System Model}
\begin{figure}[t]
	\centering
	\includegraphics[width = 0.45\textwidth]{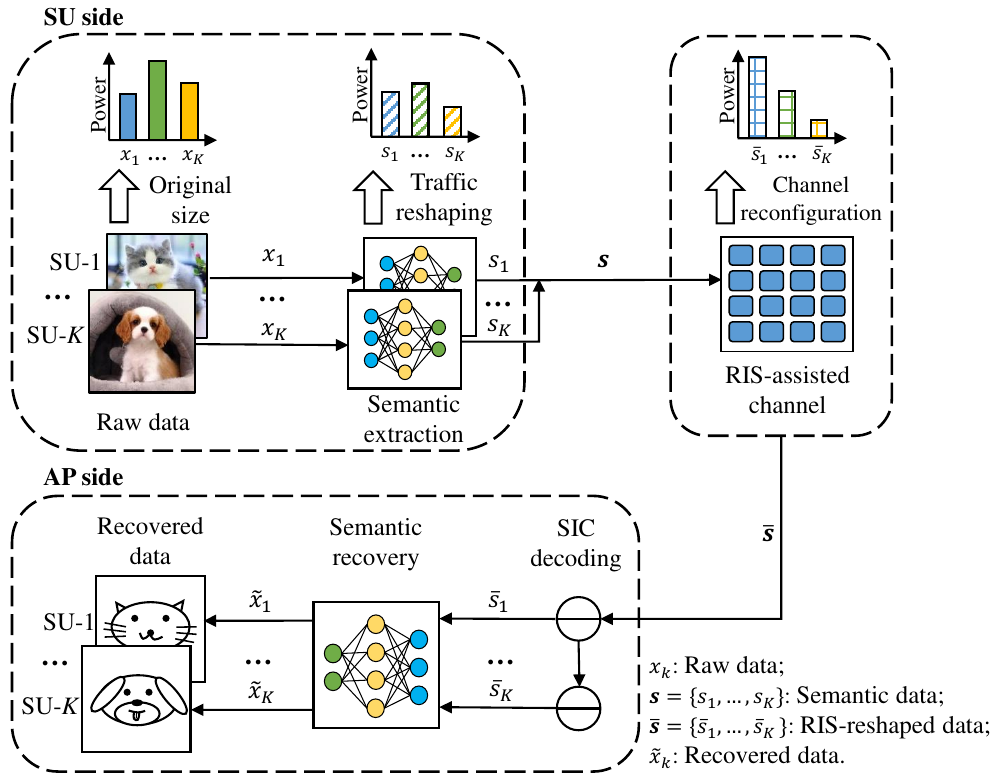}
	\caption{RIS-assisted semantic NOMA transmissions.}\label{system-model}
\vspace{-0.6cm}
\end{figure}

We consider a semantic-aware NOMA wireless network, where $K$ SUs simultaneously communicate with an AP assisted by an RIS, as shown in Fig.~\ref{system-model}. Let $\mathcal{K}=\{1,\ldots,K\}$ denote the set of the SUs and the $k$-th SU is denoted as the SU-$k$. We consider that all SUs and AP are equipped with the single antenna. We employ the NOMA method for the SUs' simultaneous access, which is expected to achieve higher spectral- and energy-efficiency compared to the conventional orthogonal multiple
access (OMA) methods~\cite{Ding-2022twc}. Each SU is equipped with a semantic control unit enabling it to extract semantic information from the raw data. The AP employs the SIC technique to decode each SU's semantic information from the superimposed signals one by one. After SIC decoding, the AP recovers each SU's raw data using a pre-trained semantic processor. The RIS can improve the SUs' channel conditions via its passive beamforming. Note that the semantic control and RIS's passive beamforming jointly adjust the SUs' traffic demands and channel conditions as needed, which are expected to significantly enhance the NOMA transmission efficiency.
\subsection{RIS-assisted Channel Model}
The RIS can induce passive beamforming to the wireless channel between the AP and SUs, which is expected to enhance the channel conditions significantly. We consider that the RIS has $L$ reflecting elements, denoted by $\mathcal{L}=\{1,\ldots,L\}$. The RIS's passive beamforming strategy is indicated by $\boldsymbol{\theta}=\{e^{j\phi_l}\}_{l\in\mathcal{L}}$, where $e^{j\phi_l}$ is the phase-shift induced by the RIS's $l$-th reflecting element. The wireless channels from the AP to the SU-$k$ and the RIS as well as the RIS to the SU-$k$ are denoted by $h_{a,k}$, ${\bf h}_{a,r}$, and ${\bf h}_{r,k}$, respectively. The equivalent channel $h_k$ from the AP to the SU-$k$ is denoted as follows:
\begin{equation}\label{ris-channel}
h_k = {\bf h}_{a,r}\text{diag}({\bf h}_{r,k})\boldsymbol{\theta} + h_{a,k}\triangleq{\bf H}_k\boldsymbol{\theta} + h_{a,k},~k\in\mathcal{K},
\end{equation}
where $\text{diag}({\bf h}_{r,k})$ is the diagonal matrix with the diagonal vector ${\bf h}_{r,k}$ and ${\bf H}_k \triangleq {\bf h}_{a,r}\text{diag}({\bf h}_{r,k})$. Note that the equivalent channel $h_k$ in~\eqref{ris-channel} can be controlled by designing the specific RIS's passive beamforming $\boldsymbol{\theta}$.
\subsection{Semantic NOMA Transmissions}
We consider that all SUs and the AP have the homogeneous semantic control units that can extract the semantic information at the SU side and then recover it at the AP~\cite{Cang-2023iotj}. The semantic control units can reshape the size of the semantic information by controlling the depth of the semantic extraction. Let $\rho_k\in [\rho_{\min},1]$ denote the SU-$k$'s semantic extraction factor, where $\rho_{\min}$ is the minimum extraction threshold determined by the intended recovery accuracy~\cite{Yang-2023jasc}. A smaller $\rho_k$ indicates that the SU-$k$ extracts deeper semantic features, thus reducing data size. After semantic extraction, the multiple SUs  transmit semantic information  by the NOMA method. The AP employs the SIC technique to decode all SUs' semantic information given the decoding order. We use a binary variable $\pi_{k,k'}$ to indicate the decoding order between the SU-$k$ and SU-$k'$. Let $\pi_{k,k'}=1$ indicate that the SU-$k$'s signal is decoded earlier while considering the SU-$k'$'s signal as the interference. Then, we have the following decoding constraint:
\begin{subequations}\label{decoding-relation}
\begin{align}
&\pi_{k,k'}+\pi_{k',k} = 1\text{ and }\pi_{k,k'}\in\{0,1\},\label{decoding-relation1}\\
&r_k<r_{k'}+M (1-\pi_{k,k'}),~k\neq k',\forall k,k'\in\mathcal{K}, \label{decoding-relation2}
\end{align}
\end{subequations}
where $\{r_k\}_{k\in\mathcal{K}}$ denote the priority indicators assigned to all SUs and $M$ is a sufficiently large constant used to prevent decoding loops.
Given the decoding order, the semantic capacity $S_k$ from SU-$k$ to the AP is represented as follows:
\begin{equation}\label{noma-throughtput}
S_k \!= \!\tau\log_2\!\Big(\!1+\frac{|h_k|^2p_k}{\sum\limits_{k'\neq k,k'\in\mathcal{K}}\pi_{k,k'}|h_{k'}|^2 p_{k'}+\sigma^2}\Big)\!,~k\in\mathcal{K},
\end{equation}
where $p_k$ and $\tau$ denote the SU-$k$'s transmit power and transmit duration, respectively. The background noise at the AP is represented by $\sigma^2$. We consider that the SU-$k$ has $Q_k$ amount of the raw sensing data to be transmitted to the AP and define ${\bf Q}=\{Q_k\}_{k\in\mathcal{K}}$. After semantic extraction, each SU's semantic capacity should satisfy $S_k\ge\rho_kQ_k$. Meanwhile, each SU's semantic capacity should ensure its successful decoding in the SIC processing. To sum up, we have the traffic demand $c(\rho_k)$ for each SU as follows:
\begin{equation}\label{su-qos}
S_k\geq \max\{S_{\min},\rho_kQ_{k}\} \triangleq c(\rho_k),~k\in\mathcal{K},
\end{equation}
where $S_{\min}$ is minimum capacity requirement determined by the signal-to-interference-plus-noise
ratio (SINR) required for decoding.
Typically, with fixed passive beamforming strategy $\boldsymbol{\theta}$, the decoding order can be determined by all SUs' channel gains~\cite{Zheng-2020cl}. However, this method may not effectively meet each SU's $\rho_k Q_k$ in~\eqref{su-qos}, which depends on the SUs' raw traffic and the semantic extraction factor. From this point, the semantic extraction factor also needs to be a key aspect in designing the NOMA decoding order.
\subsection{Energy Budget for Semantic Control and Transmission}
The system's overall energy consumption depends on both semantic control and transmission. The energy consumption in semantic control consists of both the SUs' semantic extraction and the AP's semantic recovery. Let $W_{e,k} = a Q_k/\rho^{\alpha_e}_k$ denote the additional computation load introduced by the SU-$k$'s semantic extraction, where $a>0$ and $\alpha_e>1$ denote the semantic extraction coefficients of the SUs' semantic processors. Similarly, the extracted semantic information recovered at the AP also causes the computation load $W_{r,k} = b Q_k/\rho^{\alpha_r}_k$, where $b>0$ and $\alpha_r$ is the semantic recovery coefficients~\cite{Cang-2023iotj}. Note that the $W_{e,k}$ and $W_{r,k}$ can be also modeled by the other cost functions, which can be estimated by the specific semantic model. Thus, the SU-$k$'s energy consumption caused by the semantic control $E_{s,k}$ is represented as follows:
\begin{equation}\label{sem-energy_consumption}
E_{s,k}=\kappa (f_k^2 W_{e,k} + g^2 W_{r,k}),~k\in\mathcal{K},
\end{equation}
where $f_k$ and $g$ are the computation capacities of SU-$k$ and the AP, respectively, and $\kappa$ is the energy efficiency coefficient~\cite{Bai-2021twc}.

The energy consumption in the SU-$k$'s transmission is characterized by $E_{t,k} = \tau p_k$. Combining the energy consumption from both semantic control and transmission, the overall energy consumption $E_o$ is represented as follows:
\begin{equation}\label{overall-energy-consumption}
E_{o}=\sum_{k\in\mathcal{K}}(E_{s,k}+E_{t,k}).
\end{equation}
Note that we cannot simply increase the depth of semantic extraction to reduce the energy consumption of the transmissions, as this requires more computational resources and increases semantic control energy consumption for both the SUs and the AP. As such, it is crucial to manage the trade-off between $E_{s,k}$ and $E_{t,k}$.
\vspace{-0.2cm}
\section{Energy Minimization for RIS-assisted Semantic NOMA Transmissions}
We aim to minimize the system's energy consumption of both the semantic control and transmission. Note that different decoding orders place SUs to different interference conditions and thus change their semantic capacities. From an energy saving perspective, the SUs' semantic capacities should meet their traffic demands but do not have to be very high. Thus, it is important to design the SUs' decoding order to match with their traffic demands. Additionally, the decoding order design should also consider the SUs' channel conditions. The RIS can reconfigure the SUs' channel conditions, while semantic control can reshape the SUs' traffic demands, implying that both aspects can be jointly optimized for enhancing the SUs' NOMA transmissions. To this end, we minimize the systems's overall energy by jointly optimizing the SUs' semantic extraction factor $\boldsymbol{\rho} = \{\rho_k\}_{k\in\mathcal{K}}$ and transmit power ${\bf p}=\{p_k\}_{k\in\mathcal {K}}$, the decoding order ${\boldsymbol \pi}=\{\pi_{k,k'}\}_{k,k'\in\mathcal{K}}$, and the RIS' passive beamforming strategy $\boldsymbol{\theta}$, as follows:
\begin{subequations}\label{energy-minimization}
\begin{align}
\min_{\boldsymbol{\rho},{\bf p},{\boldsymbol \pi},\boldsymbol{\theta}}&~E_{o}, \\
\mathrm {s.t.}
~&~ \eqref{ris-channel}-\eqref{overall-energy-consumption},\\
~&~ \bm{\theta}\in(0,2\pi)^L,0\le p_k\le p_{\text{max}},\rho_{\min}\le\rho_k\le1,\label{boundary-constraint}
\end{align}
\end{subequations}
where $p_{\max}$ is the SUs' maximum transmit power.
Problem~\eqref{energy-minimization} is intractable due to the integer variables and strong coupling among the control variables. We decompose the original problem~\eqref{energy-minimization} into two subproblems, and  focus on solving the control variables in each subproblem with the reduced dimensionality.
\vspace{-0.3cm}
\subsection{Penalty Relaxation for NOMA Decoding Order}
We first optimize the NOMA decoding order by fixing the other control variables. Then, the optimization of ${\boldsymbol \pi}$ becomes a feasibility check problem. A straightforward idea is that we can allocate a more favorable decoding order to the SU with heavier semantic traffic to improve the energy efficiency. This allows us to transform the feasibility check problem into a max-min problem by introducing the non-negative auxiliary variable ${\boldsymbol \ell} =\{\ell_k\}_{k\in\mathcal{K}}$, as follows:
\begin{subequations}\label{subproblem-decoding}
\begin{align}
\max_{{\boldsymbol \pi},{\boldsymbol \ell}}~\min_{k\in\mathcal{K}}&~\ell_k, \label{obj-decoding}\\
\mathrm {s.t.}
~&~ \eqref{ris-channel}-\eqref{decoding-relation},\\
~&\!\!\!\!\!\!\sum\limits_{k'\neq k,k'\in\mathcal{K}}\!\!\!\!\!\!\!\pi_{k,k'}|h_{k'}|^2 p_{k'}+\sigma^2+\ell_k\le\chi_k,\forall k\in\mathcal{K}\label{interference-relax},
\end{align}
\end{subequations}
where $\chi_k\triangleq\frac{|h_k|^2p_k}{2^{c(\rho_k)/\tau}-1}$ and the auxiliary variable $\ell_k$ can be explained by how much additional interference the SU-$k$ can tolerate. Then, the integer constraint~\eqref{decoding-relation1} is relaxed to its continuous form as follows:
\begin{subequations}\label{continuous-relaxation}
\begin{align}
&\pi_{k,k'}-\pi_{k,k'}^2 = 0 \text{ and }\pi_{k,k'}\!+\!\pi_{k',k} - 1 =0,   \label{binary}    \\
&0 \leq \pi_{k,k'} \!\leq\! 1,~k\neq k',\forall k,k'\in\mathcal{K}.\label{plus-one}
\end{align}
\end{subequations}
Since constraint~\eqref{binary} is hard to process directly, we approximate it by introducing a penalty parameter $\zeta$ as well as the auxiliary variables ${\boldsymbol\epsilon}=\{\epsilon _1,\epsilon_2\}$ and $\upsilon$. Hence, problem~\eqref{subproblem-decoding} is reformulated as follows:
\begin{subequations}\label{penalty-subproblem}
\begin{align}
\max_{{\boldsymbol \pi},{\boldsymbol \ell},{\boldsymbol\epsilon},\upsilon}&~\upsilon -\zeta(\epsilon_1+\epsilon_2), \\
\mathrm {s.t.}
~&~ \eqref{ris-channel},\eqref{decoding-relation2}, \eqref{interference-relax}, \text{ and }\eqref{plus-one},\\
~&\sum_{k\neq k',k\in\mathcal{K}}\sum_{k\neq k',k'\in\mathcal{K}} \pi_{k,k'}-\pi_{k,k'}^2\le \epsilon_1,\label{epsilon-1}\\
~&\sum_{k\neq k',k\in\mathcal{K}}\sum_{k\neq k',k'\in\mathcal{K}}\pi_{k,k'}\!+\!\pi_{k',k} - 1\le \epsilon_2,\label{epsilon-2}\\
~&\ell_k\ge\upsilon,~\forall k\in\mathcal{K}.
\end{align}
\end{subequations}
Constraint~\eqref{epsilon-1} can be linearly approximated by the first-order Taylor expansion at given points $\pi_{k,k',0}$, as follows:
\begin{equation}\label{pi-taylor-expansion}
\sum_{k\neq k',k\in\mathcal{K}}\sum_{k\neq k',k'\in\mathcal{K}}\pi_{k,k'}+\pi_{k,k',0}^2-2\pi_{k,k'}\pi_{k,k',0}\le\epsilon_1.
\end{equation}
By substituting the linear approximation~\eqref{pi-taylor-expansion} into problem~\eqref{penalty-subproblem}, it becomes a convex optimization problem and can be directly solved by standard convex solvers.
\begin{algorithm}[t]
	\caption{Joint Optimization for NOMA Decoding Order, Semantic Control, and Passive Beamforming.}\label{alg-jsap}
	\begin{algorithmic}[1]
        \State Initialize the iterative index $i=0$, parameters $\zeta^{(i)}$ and $\omega^{(i)}$, as well as decoding order ${\boldsymbol \pi}^{(i)}$. Set the overall energy consumption $E_{o}^{(i)}$ and convergence threshold $\eta>0$.
        \State \textbf{repeat}
        \State \hspace{3mm}$i = i+1$
        \State \hspace{3mm}Update ${\boldsymbol \pi}^{(i)}$ by solving subproblem~\eqref{penalty-subproblem}
        \State \hspace{3mm}Update $y_k^*$ by the closed-form solution in~\eqref{obtain-y}
        \State \hspace{3mm}Update ($\boldsymbol{\rho}^{(i)},{\bf z}^{(i)},{\bf{V}}^{(i)}$) by solving subproblem~\eqref{power-RIS-final-problem}
        \State \hspace{3mm}Update ${\bf{p}}^{(i)}$ by computing $p_k={z_k^{(i)}}^2/\text{Tr}({\bf G}_k{\bf{V}}^{(i)})$
        \State \hspace{3mm}Update $\boldsymbol{\theta}^{(i)}$ by the eigenvector decomposition of ${\bf{V}}^{(i)}$
        \State \hspace{3mm}Update $E_{o}^{(i)}$ by computing~\eqref{overall-energy-consumption}
        \State \textbf{until} $E_{o}^{(i)}-E_{o}^{(i-1)}\leq\eta$
	\end{algorithmic}
\end{algorithm}
\subsection{Optimizing for Semantic Control and Beamforming}
The RIS's passive beamforming can reconfigure the channel conditions while the semantic control can reshape the SUs' traffic demands over the wireless channels. Thus, we jointly optimize the semantic control $\{\boldsymbol{\rho},{\bf p}\}$ and RIS's passive beamforming $\boldsymbol{\theta}$ to reduce the overall energy consumption. We employ the QT technique~\cite{Shen-2018TSP} to simplify the fractional form of SINR in~\eqref{noma-throughtput} by introducing an auxiliary variable ${\bf y}=\{ y_k\}_{k\in\mathcal{K}}$ as follows:
\begin{equation}\label{SINR-QT}
\!\!\gamma_k \!= \!2y_k\!\sqrt{\text{Tr}({\bf G}_k{\bf V})p_k}\!-\!y_k^2(\!\!\!\!\!\!\!\sum\limits_{k'\neq k,k'\in\mathcal{K}}\!\!\!\!\!\!\!\pi_{k,k'}\text{Tr}({\bf G}_{k'}{\bf V})p_{k'}\!+\sigma^2),
\end{equation}
where the auxiliary matrices in constraint~\eqref{SINR-QT} are defined by $\{{\bf G}_k = [{\bf H}_k, h_{a,k}]^H[{\bf H}_k, h_{a,k}]\}_{k\in\mathcal{K}}$ and ${\bf V} = [{\boldsymbol \theta};1][{\boldsymbol \theta};1]^H$, respectively.
Thus, we reformulate problem~\eqref{energy-minimization}  as follows:
\begin{subequations}\label{power-RIS-subproblem}
\begin{align}
\min_{{\bf V} ,\boldsymbol{\rho},{\bf p}}&~E_o,\\
\mathrm {s.t.}
~&~\eqref{ris-channel},\eqref{sem-energy_consumption}-\eqref{overall-energy-consumption},\eqref{boundary-constraint},\text{ and }\eqref{SINR-QT},\\
~&~ \tau\log_2\big(1+\max_{y_k}\gamma_k\big)\ge c(\rho_k),~\forall k \in\mathcal{K},\label{re-sinr}\\
~&~{\bf V}\succeq0\text{ and }{\bf V}_{l,l}=1,~\forall l\in\{1,\ldots,L+1\},\label{semi-positive}\\
~&~\text{Rank}({\bf V})=1,\label{rank-one}
\end{align}
\end{subequations}
where $c(\rho_k)$ is linearly related to $\rho_k$, defined in~\eqref{su-qos} and ${\bf V}_{l,l}$ denotes the $l$-th  diagonal element of ${\bf V}$. Note that constraint~\eqref{re-sinr} is equivalent to constraint~\eqref{su-qos} when $y_k=\arg\max_{y_k}\gamma_k$. Note that $\gamma_k$ is concave with respect to $y_k$ in~\eqref{SINR-QT}. We can determine the optimal $y_k^*$ by setting its first derivative to zero as follows:
\begin{equation}\label{obtain-y}
y_k^*= \frac{\sqrt{\text{Tr}({\bf G}_k{\bf V})p_k}}{\!\!\!\!\!\!\sum\limits_{k'\neq k,k'\in\mathcal{K}}\!\!\!\!\!\!\!\pi_{k,k'}\text{Tr}({\bf G}_{k'}{\bf V})p_{k'}+\sigma^2},~k\in\mathcal{K}.
\end{equation}
Given $y_k^*$,
subproblem~\eqref{power-RIS-subproblem} remains intractable due to the coupling between control variables $p_k$ and ${\bf V}$ as shown in~\eqref{SINR-QT}. We introduce an auxiliary variable ${\bf z}=\{z_k\ge0\}_{k\in\mathcal{K}}$ and define $z_k^2=p_k\text{Tr}({\bf G}_k{\bf V})$. By substituting ${\bf z}$ into~\eqref{SINR-QT} and introducing an auxiliary variable $\boldsymbol{\widehat{\gamma}}=\{\widehat{\gamma}_k\}_{k\in\mathcal{K}}$, we approximate constraint~\eqref{SINR-QT} to a convex form as follows:
\begin{equation}\label{SINR-QT-substitute}
\widehat{\gamma}_k \le 2y_k^*z_k\!-\!{y_k^*}^2(\!\!\!\!\!\sum\limits_{k'\neq k,k'\in\mathcal{K}}\!\!\!\!\!\!\pi_{k,k'}z^2_{k'}+\sigma^2),~\forall k\in\mathcal{K}.
\end{equation}

To tackle constraint~\eqref{rank-one}, we apply the sequential rank-one constraint relaxation (SROCR) method~\cite{Cao-2017european} by introducing a relaxation variable $\omega\in[0,1]$ as follows:
\begin{equation}\label{reformulate-rank-one}
{\bf e}^H{\bf V}{\bf e}\ge \omega\text{Tr}({\bf V}),
\end{equation}
where ${\bf e}$ is the eigenvector to the largest eigenvalue of ${\bf V}$. The main idea of SROCR method is to gradually tighten constraint~\eqref{reformulate-rank-one} to approximate~\eqref{rank-one} by sequentially increasing $\omega$ from $0$ to $1$. The vector ${\bf e}$ is updated from ${\bf V}$ in last iteration. Substituting constraints~\eqref{SINR-QT-substitute} and~\eqref{reformulate-rank-one} into problem~\eqref{power-RIS-subproblem}, we update the problem as follows:
\begin{subequations}\label{power-RIS-final-problem}
\begin{align}
\min_{{\bf V} ,\boldsymbol{\rho},{\bf z},\boldsymbol{\widehat{\gamma}}}&~\sum_{k\in\mathcal{K}}\frac{\kappa af_k^2 Q_k}{\rho^{\alpha_e}_k}+ \frac{\kappa b g^2 Q_k}{\rho^{\alpha_r}_k}+\frac{\tau z_k^2}{\text{Tr}({\bf G}_k{\bf V})},\\
\mathrm {s.t.}
~&~\eqref{ris-channel},\eqref{boundary-constraint},\eqref{semi-positive}, \text{ and }\eqref{SINR-QT-substitute}-\eqref{reformulate-rank-one}, \\
~&~ \tau\log_2\left(1+\widehat{\gamma}_k\right)\ge c(\rho_k),~\forall k \in\mathcal{K}.\label{re-sinr-renew}
\end{align}
\end{subequations}
Note that $z_k^2/\text{Tr}({\bf G}_k{\bf V})$ is joint convex with respect to $z_k$ and ${\bf V}$. As such, problem~\eqref{power-RIS-final-problem} becomes a standard convex optimization problem. After obtaining the solution in each iteration, the SUs' transmit power ${\bf p}$ is determined by $p_k=z_k^2/\text{Tr}({\bf G}_k{\bf V})$ and the RIS's passive beamforming $\boldsymbol{\theta}$ is obtained by the eigenvector decomposition of ${\bf V}$.

The detailed solution procedure is summarized in Algorithm~\ref{alg-jsap}. The subproblems alternatively iterate until the overall energy consumption $E_{o}$ converges. To ensure convergence, the value of $E_{o}$ needs to decrease monotonically after solving each subproblem. However, the penalty method in~\eqref{penalty-subproblem} and the SROCR method in~\eqref{power-RIS-final-problem} cannot guarantee a decreasing $E_{o}$ with every iteration. To address this issue, we compare $E_{o}$ across adjacent iterations and retain the control variables with a lower $E_{o}$. The computational complexity of Algorithm~\ref{alg-jsap} is estimated by $\mathcal{O}\big((K)^{7}+(3K+L+1)^{3.5}\big)I_{\max}$~\cite{Luo-2020spm}, where $I_{\max}$ is the required iteration number.
\vspace{-0.2cm}
\section{Numerical Results}
In this section, we present the numerical results to evaluate the energy saving performance of the NOMA transmissions by the proposed joint traffic reshaping and channel reconfiguration scheme (denoted as JTAC scheme). We compare the JTAC scheme with three benchmark schemes, i.e., the Fixed-Phase, the Fixed-Semantic, and the Channel-Decoding schemes. The Fixed-Phase and Fixed-Semantic schemes represent that the RIS applies fixed phase shifts to all reflecting elements and the SUs are allocated with the fixed semantic extraction factors, respectively. The Channel-Decoding scheme determines the SUs' decoding order by sorting their channel gains in descending order, such as in~\cite{Wang-2022twc}. The channel model is assumed to follow the long-distance propagation, where the path loss at the reference distance is set as $L_0=30$ dB. We consider that the AP is located at the coordinate origin and the RIS is initially located at $(5,0)$ m. We set three SUs placed at $(6,2)$  m, $(8,1.5)$  m, and $(8,2)$  m, respectively. Similar to the settings in~\cite{Cang-2023iotj}, the default parameters are set as follows: $p_{\max} = 40$ dBm, $\rho_{\min} = 0.2$, $Q_k=10$ Kbits, $\kappa = 10^{-21}$, $a =100$, $b = 200$, $\alpha_e =4$, $\alpha_r =1$, $f_k = 5\times10^{8}$ cycles/s, $g=10^9$ cycles/s, $\eta = 10^{-3}$, and $\sigma^2 = -80$ dBm.

\begin{figure}[t]
	\centering
    \subfloat[Convergence comparison.]{\includegraphics[width=0.23\textwidth]{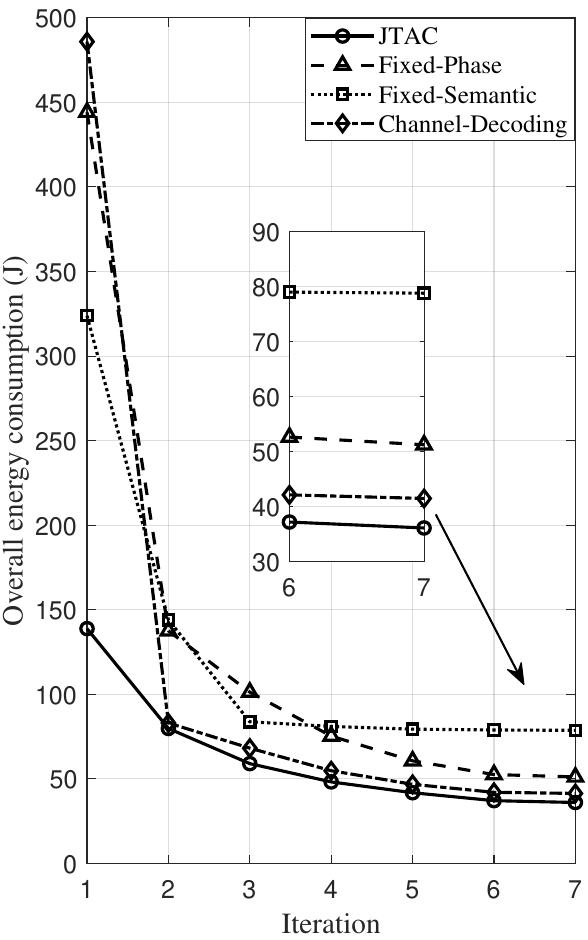}}
	\subfloat[Different locations of the RIS.]{\includegraphics[width=0.23\textwidth]{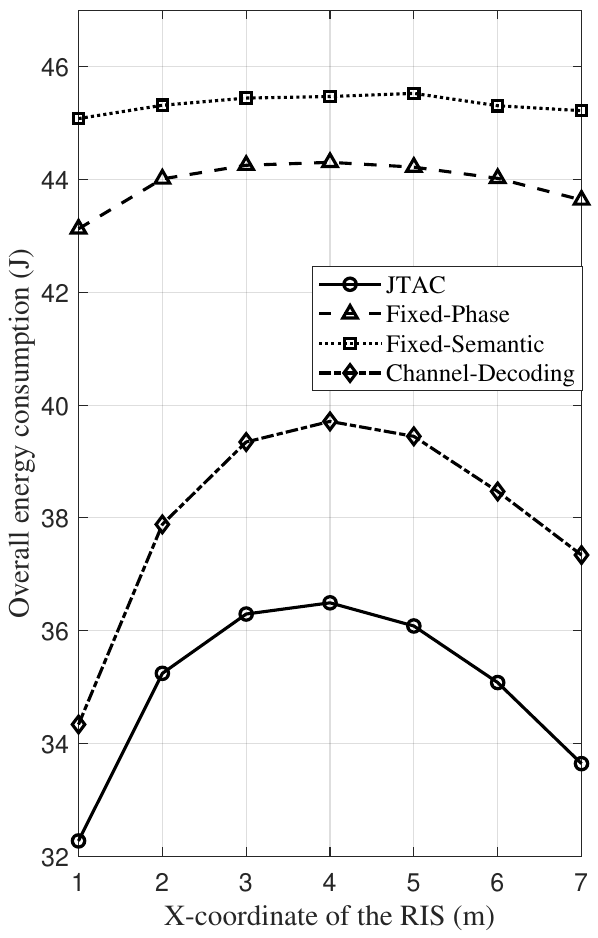}}
	\caption{Energy saving performance under different schemes.}\label{energy_consumption}
\vspace{-0.7cm}
\end{figure}

Figure~\ref{energy_consumption}(a) validates the convergence performance of the different schemes. It is observed that all schemes converge with the significant reductions in the overall energy consumption as the iterations proceed. The JTAC scheme shows the best energy saving performance compared to the benchmark schemes. This is because the joint semantic control and RIS's passive beamforming can provide greater flexibility for the SUs' NOMA transmissions, which makes it easier to satisfy all SUs' traffic demands and thus reduce the overall energy consumption significantly. It is also note that the Channel-Decoding scheme outperforms the other two benchmark methods. This indicates that the joint semantic control and RIS's passive beamforming can help the SUs efficiently adapt their current decoding order even if it is not ideal. All schemes show the fast and stable convergence, which implies the practical applicability of Algorithm~\ref{alg-jsap}.

The different RIS's locations provide the SUs with different reflection capabilities, which directly affect their NOMA transmission efficiencies. Thus, we explore the impact of the RIS's location on all schemes' energy saving performance as shown in Fig.~\ref{energy_consumption}(b). We vary the RIS's location from $(1,0)$ m to $(7,0)$ m. It is seen that the overall energy consumption first increases as the RIS moves away from the AP. This is due to the decreasing reflection performance of the AP-RIS links, leading to deteriorating channel conditions. When the RIS's x-coordinate exceeds $4$ m and gets closer to SUs, the RIS's reflection becomes more dominant and effective. As such, the overall energy consumption begins to decrease. This implies that the RIS can be deployed either close to  SUs or the AP to better enhance  reflection efficiency. The JTAC scheme shows superior energy saving performance compared to the benchmarks, particularly when the RIS is deployed near SUs and the AP. This validates the effectiveness of the JTAC scheme under various channel conditions.

\begin{figure}[t]
	\centering
    \subfloat[Different sizes of the RIS.]{\includegraphics[width=0.23\textwidth]{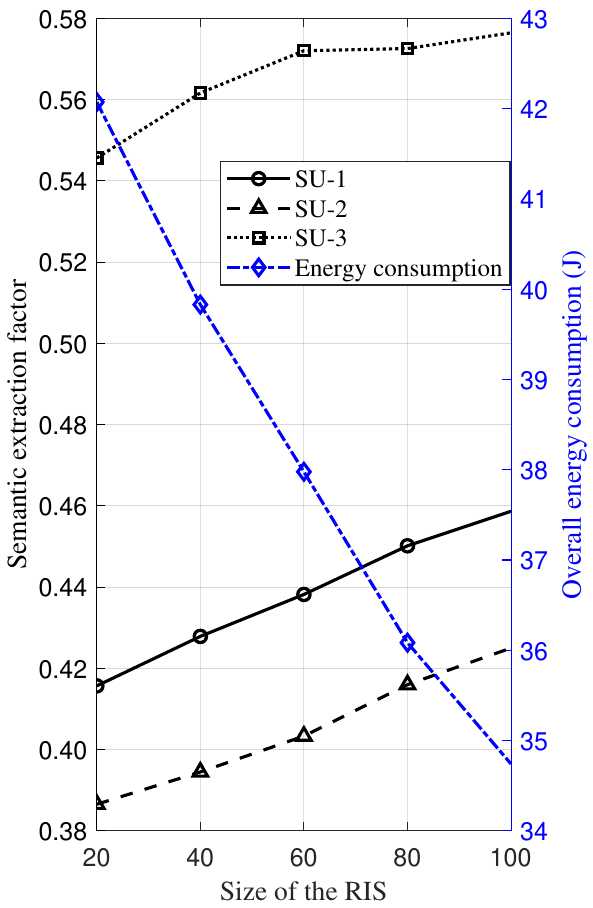}}
	\subfloat[Different raw data sizes.]{\includegraphics[width=0.23\textwidth]{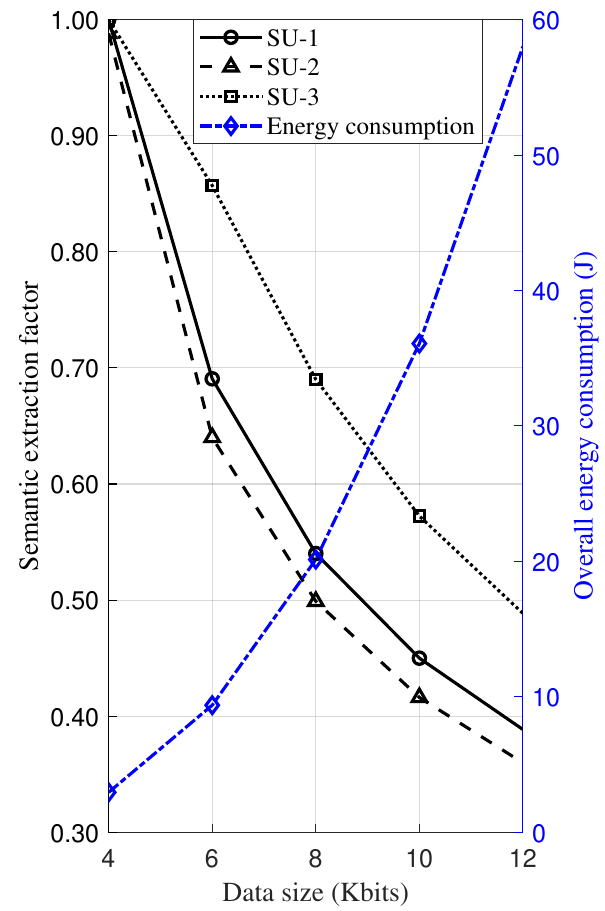}}
	\caption{Semantic control strategy under different parameters.}\label{parameter}
\vspace{-0.7cm}
\end{figure}

We have demonstrated the superior performance of the JTAC scheme for the SUs' NOMA transmissions. In the following, we delve into the JTAC scheme and evaluate it with different simulation parameters. Figure~\ref{parameter}(a) investigates how the JTAC scheme performs as we change the element number of the RIS from $20$ to $100$. It is evident that the overall energy consumption decreases significantly as increasing the RIS's size $L$. This is because the RIS with a larger size can offer greater channel enhancement for the SUs' NOMA transmissions. Besides, we observe that the semantic extraction factor $\boldsymbol{\rho}$ is proportional to the RIS's size. The reason is that more preferable channel conditions lead to stronger transmission capacities for the SUs, which reduces the energy consumption during the transmissions. This motivates the SUs to transmit larger-size semantic information with lager $\boldsymbol{\rho}$.

Figure~\ref{parameter}(b) examines the energy saving performance with the varying size of the SUs' raw data. For simplicity, we consider that all SUs have the same size ${\bf Q}$ and increase it from $4$ Kbits to $12$ Kbits. It is observed that the overall energy consumption increases in ${\bf Q}$. This is straightforward as the SUs need more energy for both the semantic control and transmission when the raw data becomes heavier. Interestingly, we observe that the SUs can transmit raw data without any extraction when ${\bf Q}$ is low. As ${\bf Q}$ increases, the SUs tend to decrease the $\boldsymbol{\rho}$ to lighten the energy consumption of transmissions. This validates that the JTAC scheme effectively balances the energy trade-off between the semantic control and transmissions according to varying  ${\bf Q}$.
\section{Conclusions}
In this paper, we have investigated an semantic-aware and RIS-assisted NOMA wireless network. The joint traffic reshaping by the semantic control and the RIS' channel
reconfiguration is expected to bring greater degree of freedom for the SUs' information decoding. The system's overall energy consumption is minimized by jointly optimizing the SUs' decoding order and semantic control, as well as the RIS's passive beamforming. The complex problem is decomposed into two subproblems, which are solved by employing the penalty relaxation and the SROCR methods. Numerical results have validated that the JTAC scheme significantly reduces the overall energy consumption, which can be further improved by positioning the RIS closer to either the AP or the SUs. Additionally, we note that the NOMA decoding order optimization may introduce a high computational complexity $\mathcal{O}(K^7)$ as the number of the SUs increases. Thus, designing algorithms with lower complexity will be investigated in our future work.

\footnotesize
\bibliographystyle{IEEEtran}
\bibliography{reference}
\end{document}